\documentclass[prl,twocolumn,letterpaper,superscriptaddress]{revtex4-1}
\usepackage{bm,graphicx,graphics,amsmath,amssymb,bm,epsfig,color}
\usepackage{euscript,tabularx} \usepackage{longtable}

\begin{document}

\title{Perfect and robust phase-locking of a spin transfer vortex nano-oscillator \\
  to an external microwave source}

\author{A. Hamadeh}
\affiliation{Service de Physique de l'\'Etat Condens\'e (CNRS URA
  2464), CEA Saclay, 91191 Gif-sur-Yvette, France}

\author{N. Locatelli}
\affiliation{Unit\'e Mixte de Physique CNRS/Thales and Universit\'e
  Paris Sud 11, 1 av. Fresnel, 91767 Palaiseau, France}

\author{V.V. Naletov}
\affiliation{Service de Physique de l'\'Etat Condens\'e (CNRS URA
  2464), CEA Saclay, 91191 Gif-sur-Yvette, France}
\affiliation{Unit\'e Mixte de Physique CNRS/Thales and Universit\'e
  Paris Sud 11, 1 av. Fresnel, 91767 Palaiseau, France}
\affiliation{Institute of Physics, Kazan Federal University, Kazan
    420008, Russian Federation}

\author{R. Lebrun}
\affiliation{Unit\'e Mixte de Physique CNRS/Thales and Universit\'e
  Paris Sud 11, 1 av. Fresnel, 91767 Palaiseau, France}

\author{G. de Loubens}
\email{gregoire.deloubens@cea.fr}
\affiliation{Service de Physique de l'\'Etat Condens\'e (CNRS URA
  2464), CEA Saclay, 91191 Gif-sur-Yvette, France}

\author{J. Grollier}
\affiliation{Unit\'e Mixte de Physique CNRS/Thales and Universit\'e
  Paris Sud 11, 1 av. Fresnel, 91767 Palaiseau, France}

\author{O. Klein}
\affiliation{Service de Physique de l'\'Etat Condens\'e (CNRS URA
  2464), CEA Saclay, 91191 Gif-sur-Yvette, France}

\author{V. Cros}
\affiliation{Unit\'e Mixte de Physique CNRS/Thales and Universit\'e
  Paris Sud 11, 1 av. Fresnel, 91767 Palaiseau, France}

\date{\today}

\begin{abstract}

  We study the synchronization of the auto-oscillation signal
  generated by the spin transfer driven dynamics of two coupled
  vortices in a spin-valve nanopillar to an external
  source. Phase-locking to the microwave field $h_\text{rf}$ occurs in
  a range larger than 10\% of the oscillator frequency for drive
  amplitudes of only a few Oersteds. Using synchronization at the
  double frequency, the generation linewidth is found to decrease by
  more than five orders of magnitude in the phase-locked regime (down
  to 1~Hz, limited by the resolution bandwidth of the spectrum
  analyzer) in comparison to the free running regime ($140$~kHz). This
  perfect phase-locking holds for frequency detuning as large as
  2~MHz, which proves its robustness. We also analyze how the free
  running spectral linewidth impacts the main characteristics of the
  synchronization regime.

\end{abstract}

\maketitle

Spin transfer nano-oscillators (STNOs) are nanoscale microwave
generators \cite{kiselev03,rippard04} which have become very
attractive due to their wide range of potential applications
(frequency generation \cite{houssameddine07,bonetti09} and detection
\cite{tulapurkar05,zhu12}, signal processing
\cite{muduli10,pogoryelov11}, dynamic recording
\cite{mizushima10,zhu10}). The transfer of angular momentum from a
spin-polarized current to a ferromagnetic layer can excite the
gyrotropic mode of a magnetic vortex \cite{pribiag07,mistral08} having
typical frequency between 20~MHz and 2~GHz
\cite{guslienko08}. Vortex-based STNOs are very promising due to their
narrow generation linewidth (about 1~MHz) and potentially high output
power \cite{dussaux10}. Recently, we have proposed a way to minimize
even more the auto-oscillation linewidth by operating a STNO based on
two coupled vortices in a spin-valve nanopillar, which can yield
highly coherent signals ($Q>15000$) with linewidths under 50~kHz at
room temperature and near zero magnetic field \cite{locatelli11}.

Synchronization to an external periodic signal and mutual
phase-locking of several STNOs have been proposed as means to increase
the emitted power and reduce the phase noise of STNOs
\cite{slavin09}. It has also been suggested that synchronized arrays
of STNOs could be operated as associative memories
\cite{shibata12}. So far, mutual phase-locking has been achieved using
spin wave coupling between nanocontacts \cite{kaka05,mancoff05,sani13}
and 2D arrays of vortices and anti-vortices \cite{ruotolo09}. It is
also predicted to occur using the common microwave current emitted
\cite{slavin05a, grollier06} or the dipolar interaction between
adjacent STNOs \cite{belanovsky12,erokhin13}. To demonstrate the
efficiency of these two types of coupling, synchronization to an
external microwave current passing through the device
\cite{rippard05,georges08,quinsat11,dussaux11} or to a microwave field
produced by an external antenna \cite{urazhdin10,hamadeh12} have been
studied.

Two key characteristics to analyze the quality of the synchronization
are the locking range and the generation linewidth in the phase-locked
regime, which are respectively related to the coupling efficiency and
the response to noise of the oscillator. In a single vortex-based
tunneling magnetoresistance (TMR) device, it was shown that using an
external microwave current, the locking range could reach up to one
third of the oscillator frequency, and the linewidth be reduced by 3
orders of magnitude, from a few MHz down to 3~kHz \cite{dussaux11}. In
this letter, we demonstrate perfect and robust synchronization of the
microwave signal generated by the dynamics of two coupled vortices in
a spin-valve nanopillar to an external microwave field
$h_\text{rf}$. The linewidth measured in the phase-locked regime is
indeed limited by the minimal resolution bandwidth (RBW) of the
spectrum analyzer, which is 1~Hz. We observe such outstanding
characteristics even for frequency detunings larger than ten times the
free running linewidth (140~kHz).

\begin{figure}
  \includegraphics[width=\columnwidth]{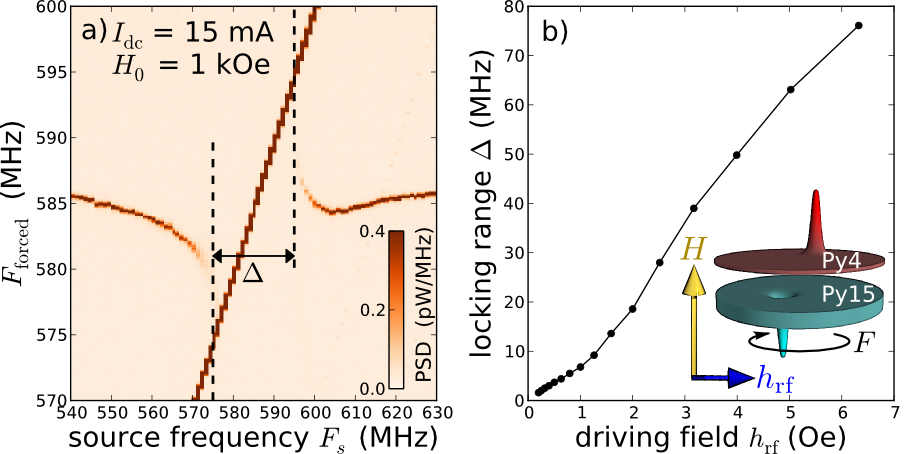}
  \caption{(color online) (a) Power spectrum map of the STNO at
    $I_\text{dc}=15$~mA and $H_0=1$~kOe vs. the frequency $F_s$ of the
    external microwave field $h_\text{rf}=2$~Oe. (b) Locking range
    $\Delta$ as a function of the drive amplitude.}
  \label{fig:1}
\end{figure}

The studied STNO is a circular nanopillar of diameter 250~nm patterned
from a (Cu60$|$Py15 $|$Cu10$|$Py4$|$Au25) stack, where thicknesses are
in nm and Py=Ni$_{80}$Fe$_{20}$. An insulating resist is deposited
onto the STNO device and an external antenna is patterned on top to
generate a spatially uniform microwave magnetic field $h_\text{rf}$
oriented in the plane of the magnetic layers \cite{naletov11}. By
injecting a current $I_\text{dc}>0$ through the STNO (electrons
flowing from the thick to the thin Py layer), a vortex with chirality
parallel to the orthoradial Oersted field is stabilized in each of the
Py layers \cite{locatelli11,sluka12}. A magnetic field $H$ is applied
perpendicularly to the sample plane and the vortex core polarities are
set to be anti-parallel (see inset of Fig.\ref{fig:1}b). For
$I_\text{dc}\gtrsim 10$~mA, a narrow microwave emission peak
corresponding to the spin transfer driven dynamics of the two coupled
vortices is detected on the spectrum analyzer. At fixed $I_\text{dc}$,
the microwave characteristics of this auto-oscillation peak (frequency
and linewidth) can be tuned by varying $H$ \cite{hamadeh13}. In this
study, all measurements are carried out at room temperature.


The perpendicular field is first set to $H_0 = 1$~kOe and the dc
current fixed to $I_\text{dc}=15$~mA. Under these bias conditions, the
oscillator frequency is $F_0=586$~MHz and the generation linewidth
$\Delta F_0=142$~kHz. In Fig.\ref{fig:1}a, we present a map of the
power density when the frequency $F_s$ of the external microwave field
is swept from 540~MHz to 630~MHz at constant drive amplitude
$h_\text{rf}=2$~Oe \footnote{The output power from the synthetizer
  injected into the microwave antenna is set to $P=0$~dBm.}. When
$F_s$ comes closer to $F_0$, the frequency of the oscillator is pulled
towards the source frequency. When $F_s\simeq 574$~MHz, there is a
single frequency peak in the spectrum, meaning that the
auto-oscillation is synchronized to the external source. At this
point, it is not possible to separate the signal of the gyrotropic
oscillation and that of the source, which prevents measuring the
generation linewidth in the phase-locked regime. This situation is
observed until $F_s\simeq 597$~MHz, above which the oscillation
frequency gradually shifts back to its free running value $F_0$. The
locking range $\Delta$ measured experimentally is plotted
vs. $h_\text{rf}$ in Fig.\ref{fig:1}b. As expected \cite{slavin09}, it
increases linearly with $h_\text{rf}$ at low drive amplitude
($h_\text{rf}<1.5$~Oe). The behavior observed at larger $h_\text{rf}$
is presumably due to some nonlinearities of the system. We point out
that at $h_\text{rf}=6.3$~Oe, the locking range $\Delta=75$~MHz
corresponds to 13\% of the oscillator frequency $F_0$.


\begin{figure}
  \includegraphics[width=7.5cm]{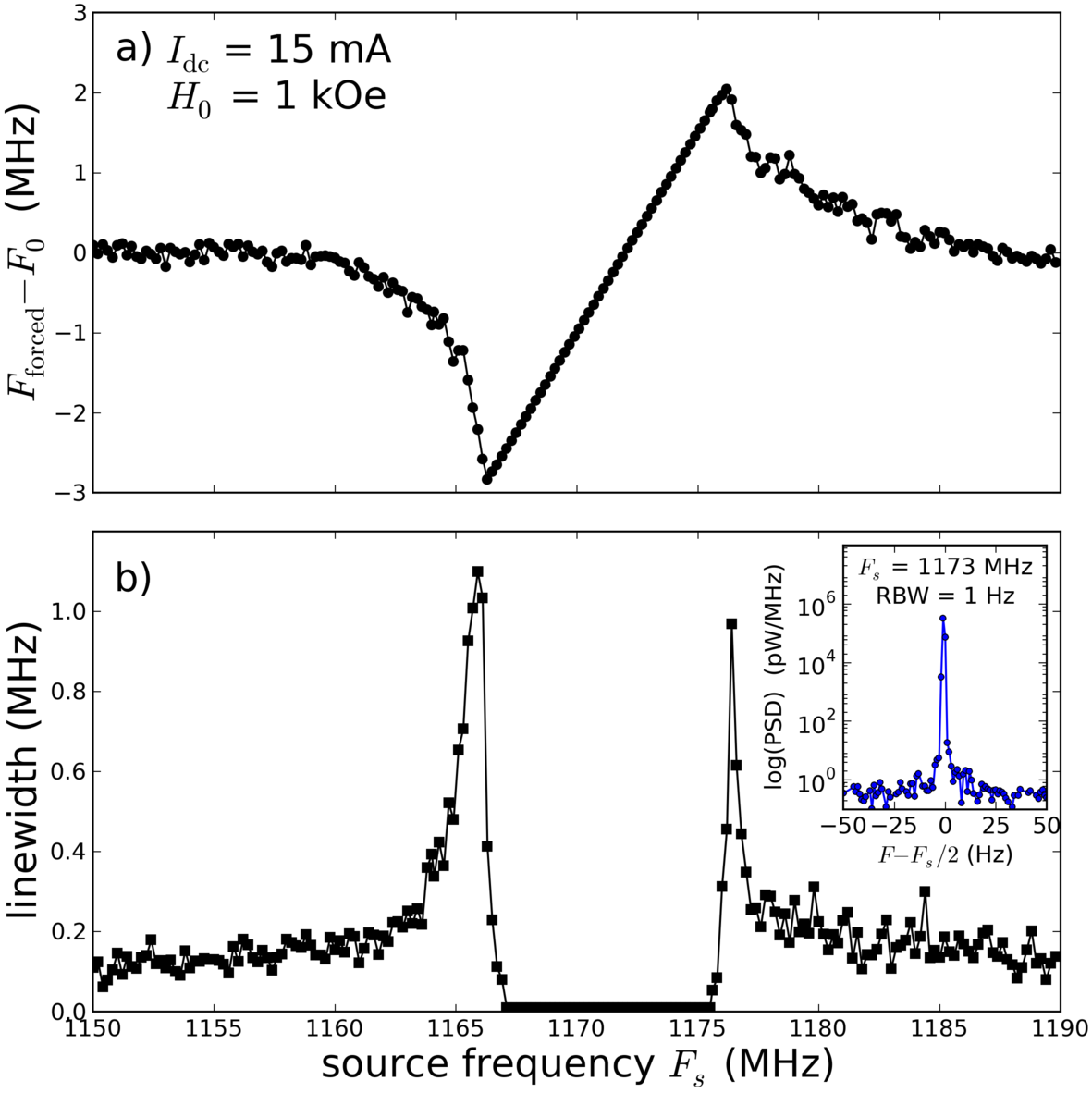}
  \caption{(color online) (a) Frequency shift $F_{\text{forced}}-F_0$
    and (b) linewidth of the generated signal as a function of the
    frequency of the source ($h_\text{rf}=6.3$~Oe), swept around
    $2F_0$. The inset displays a measurement in the locking range
    (spectrum analyzer RBW$=1$~Hz).}
  \label{fig:2}
\end{figure}


In order to measure the linewidth of the oscillator signal when its
frequency is locked, the source frequency $F_s$ is now swept around
$2F_0$. In Fig.\ref{fig:2}a, we plot the frequency shift
$F_{\text{forced}}-F_0$ of the oscillator when it is forced by the
microwave field of amplitude $h_\text{rf}=6.3$~Oe as a function of
$F_s$ varying from 1150~MHz to 1190~MHz. As in Fig.\ref{fig:1}a, we
observe the characteristic behavior of synchronization to the external
source, except that it is now at twice the oscillator frequency and
the oscillation signal is not hindered by the source signal. Hence, we
can analyze the dependence of the generation linewidth on $F_s$, which
is plotted in Fig.\ref{fig:2}b.  The striking observation is a
dramatic reduction of the generation linewidth within the locking
range. As shown in the inset of Fig.\ref{fig:2}b, the measured
linewidth is indeed limited by the 1~Hz minimal RBW of the spectrum
analyzer, \textit{i.e.}, the auto-oscillation is perfectly
phase-locked to the external source.  This corresponds to an
improvement of the signal coherency by a factor greater than $10^5$
with respect to the free running case. The increase of the generation
linewidth up to 1~MHz observed at the boundaries of the locking range
is attributed to successive synchronization-unsynchronization events
occurring at the timescale of the measurement \cite{dussaux11}.


\begin{figure}
  \includegraphics[width=7.5cm]{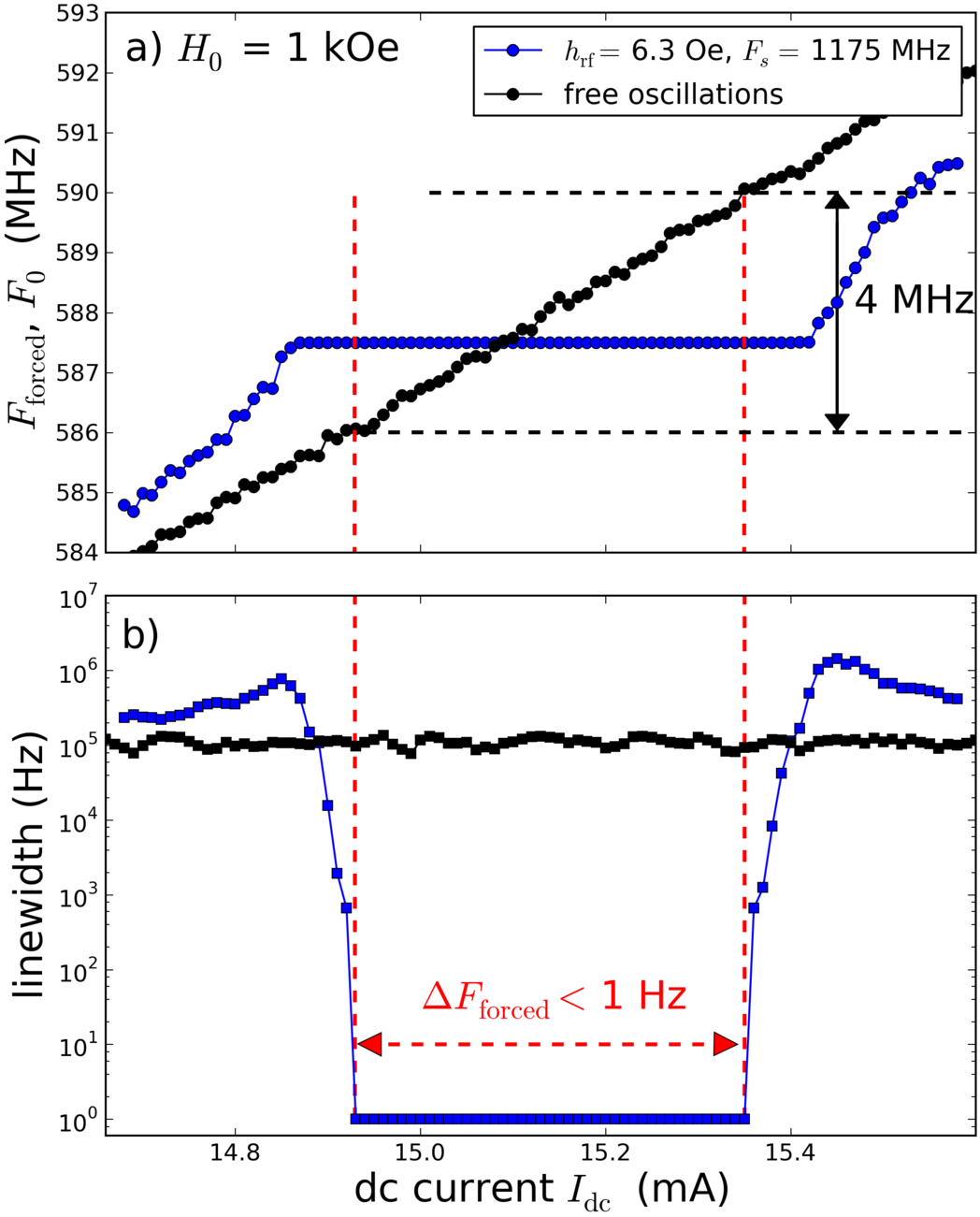}
  \caption{(color online) Current dependence of the (a) STNO frequency
    and (b) generation linewidth in the free (black dots) and forced
    regimes (blue dots).}
  \label{fig:3}
\end{figure}


To gain further insight, we investigate the robustness of this perfect
phase-locking. We now measure the auto-oscillation signal as a
function of $I_\text{dc}$, which is swept from 14.6~mA to 15.6~mA. In
the free regime (external source turned off), the generation frequency
increases linearly from 584~MHz to 592~MHz, while the linewidth is
nearly constant around $\Delta F_0=142$~kHz, as shown by the black
dots in Figs.\ref{fig:3}a and b, respectively. The tunability observed
in our vortex-based STNO, $dF_0/dI_\text{dc} \simeq 8$~MHz/mA, results
from the Oersted field created by the dc current
\cite{khvalkovskiy09}. In the forced regime with the external source
turned on at $F_s=1175$~MHz and $h_\text{rf}=6.3$~Oe (see blue dots in
Fig.\ref{fig:3}a), the auto-oscillation frequency is pulled towards
half the source frequency $F_s/2$ for $I_\text{dc}<14.9$~mA and
$I_\text{dc}>15.4$~mA, and constant and equal to $F_s/2$ in between
these boundaries, which define the locking range. The associated
decrease of the generation linewidth is spectacular, as shown by the
logarithmic scale in Fig.\ref{fig:3}b. The measured linewidth is
limited by the $\text{RBW}=1$~Hz for $14.93<I_\text{dc}<15.35$~mA,
which means that the phase-locking to the external source is perfect
within this range of current. The latter corresponds to a variation by
4~MHz of the auto-oscillation frequency in the free regime. These
features demonstrate the robustness of the synchronization observed in
our sample, as it means that even if the external source frequency
deviates from the oscillator frequency by more than ten times the free
running linewidth, perfect phase-locking can still occur.


\begin{figure}
  \includegraphics[width=7.5cm]{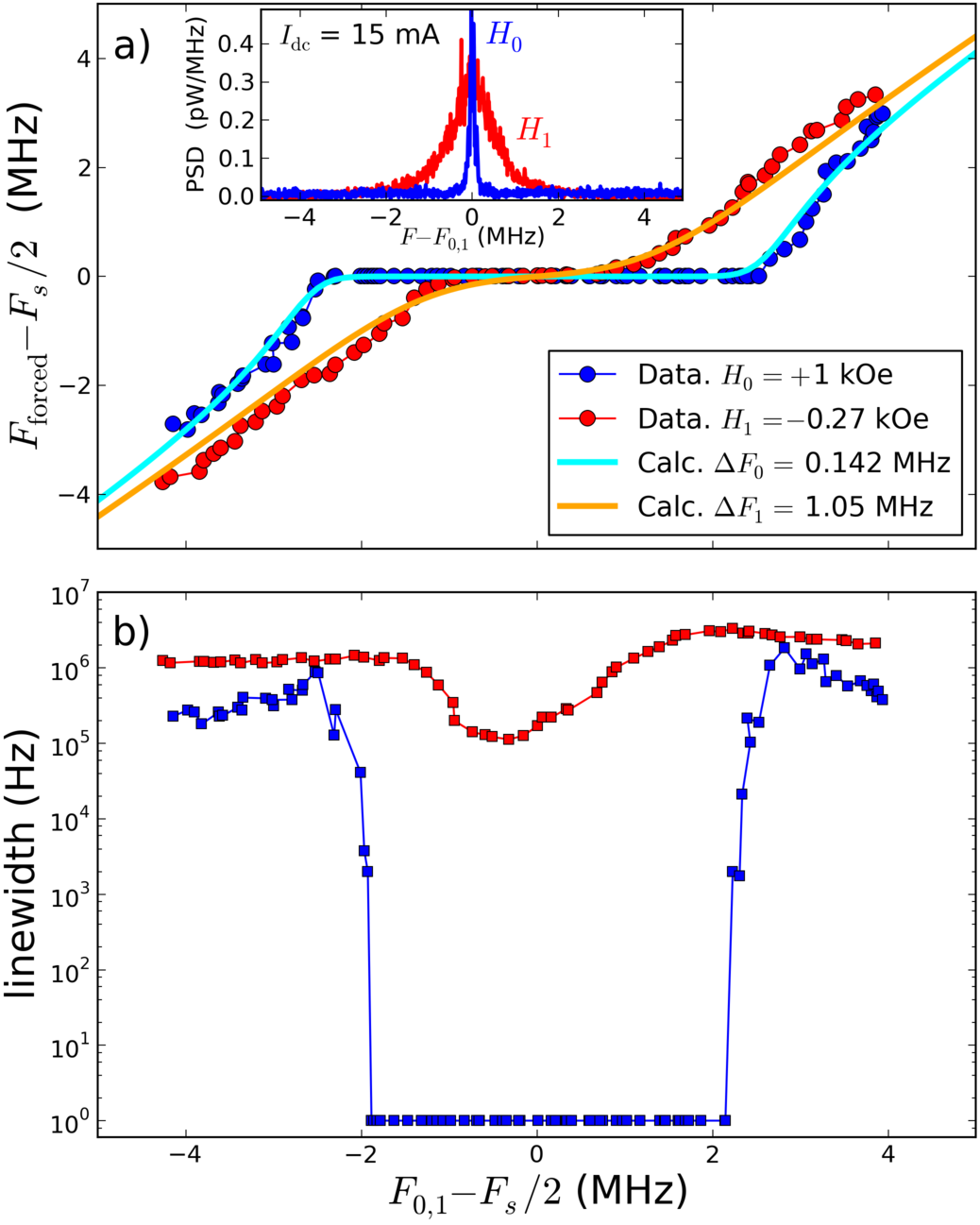}
  \caption{(a) Variation of the frequency mismatch
    $F_\text{forced}-F_s/2$ as a function of the detuning
    $F_{0,1}-F_s/2$ at $H_0 = +1$~kOe (blue dots) and $H_1 =
    -0.27$~kOe (red dots). The external source amplitude is set to
    $h_\text{rf}$ = 6.3 Oe. Continuous lines are fits using Eq.(5) of
    Ref.\cite{georges08} yielding a coupling strength
    $\varepsilon=2.5$~MHz. The inset shows the emission spectra at
    $H_0$ and $H_1$ in the free running regime. (b) Dependence of the
    emission linewidth on the frequency detuning at $H_0$ and $H_1$.}
  \label{fig:4}
\end{figure}


Another issue to investigate is the influence of fluctuations
\cite{grimaldi13} on the actual characteristics of our vortex
oscillator when it is phase-locked. To do that, we compare the
synchronization of auto-oscillation signals having different
generation linewidths. We use two different applied fields,
$H_0=+1$~kOe and $H_1=-0.27$~kOe, at which the emission frequencies at
$I_\text{dc}=15$~mA slightly differ ($F_0=586$~MHz and $F_1=684$~MHz,
respectively), and the generation linewidth varies by more than a
factor seven \footnote{This change of linewidth is due to the
  influence of a lower frequency overdamped mode \cite{hamadeh13}.},
from $\Delta F_0=142$~kHz to $\Delta F_1=1.05$~MHz (see inset of
Fig.\ref{fig:4}a). Using blue and red dots at $H_0$ and $H_1$,
respectively, we plot the experimental frequency mismatch
$F_\text{forced}-F_s/2$ (Fig.\ref{fig:4}a) and the linewidth in the
forced regime (Fig.\ref{fig:4}b) as a function of the detuning
$F_{0,1}-F_s/2$ between the natural oscillator frequency and half the
source frequency \footnote{In these measurements, $h_\text{rf}=6.3$~Oe
  and $I_\text{dc}$ is varied from 14.6~mA to 15.6~mA. At $H_0$, $F_0$
  varies from 584~MHz to 592~MHz and $F_s$ is fixed to 1175~MHz. At
  $H_1$, $F_1$ varies from 681~MHz to 689~MHz and $F_s$ is fixed to
  1370~MHz.}. The strong differences observed in the characteristics
of the synchronization at these two fields reveal the role played by
the fluctuations in the phase dynamics of STNOs. When the latter are
weak (narrower generation linewidth at $H_0$), the locking range is
large (more than 4~MHz) and the synchronized signal acquires the
spectral quality of the source (less than 1~Hz). When the noise is
larger (broader generation linewidth at $H_1$), it competes against
the coupling to the external source, which results in a smaller
apparent locking range and a poorest spectral quality of the forced
oscillation. Here, increasing the linewidth by a factor $\Delta
F_1/\Delta F_0 \simeq 7$ has a huge influence on the signal coherency
in the phase-locked regime since its improvement with respect to the
free running case drops from a factor $10^5$ to only 10. The influence
of phase fluctuations on the frequency mismatch has been modeled by
Eq.(5) of Ref.\cite{georges08} (see continuous lines in
Fig.\ref{fig:4}a). Using the measured linewidths $\Delta F_0$ and
$\Delta F_1$ in this equation, the only fitting parameter is the
coupling strength of the external microwave source to the oscillator
(equal to half the locking range in the case of zero fluctuations),
which is found to be $\varepsilon=2.5$~MHz both at $H_0$ and $H_1$.

In conclusion, we have shown that the microwave signal generated by a
STNO based on coupled vortices can be efficiently synchronized to an
external microwave field. The relative locking range indeed exceeds
10\% for small drive amplitudes ($h_\text{rf}\simeq 5$~Oe) and the
auto-oscillation signal acquires the spectral purity of the source,
corresponding to an improvement of its coherency by a factor greater
than $10^5$. Moreover, this perfect phase-locking is robust, as it
survives even when the external frequency deviates from the oscillator
frequency by more than ten times its linewidth. We believe that the
efficient synchronization of vortex-based STNOs to the microwave field
is very promising for the idea of mutually coupling such oscillators
through the dipolar interaction \cite{belanovsky13}.

This research was partly funded by the French ANR (grant SPINNOVA
ANR-11-NANO-0016) and the EU (FP7 grant MOSAIC ICT-FP7-317950).


%

\end{document}